\documentclass[aps,prd,twocolumn,groupedaddress,amssymb,eqsecnum,showpacs,
nofootinbib]{revtex4}
\usepackage{graphicx}
\usepackage{bm}

\begin{document}

\preprint{CMU-HEP-03-11}

\title{Taming the $\alpha$-vacuum}

\author{Hael Collins}
\email{hcollins@andrew.cmu.edu}
\author{R.~Holman}
\email{rh4a@andrew.cmu.edu}
\affiliation{Department of Physics, Carnegie Mellon University, 
Pittsburgh PA\ \ 15213}


\begin{abstract}
An interacting scalar field theory in de Sitter space is non-renormalizable for a generic $\alpha$-vacuum state.  This pathology arises since the usual propagator used allows for a constructive interference among propagators in loop corrections, which produces divergences that are not proportional to standard counterterms.  This interference can be avoided by defining a new propagator for the $\alpha$-vacuum based on a generalized time-ordering prescription.  The generating functional associated with this propagator contains a source that couples to the field both at a point and at its antipode.  To one loop order, we show that a set of theories with very general antipodal interactions is causal and renormalizable.  
\end{abstract}


\maketitle

\section{Introduction}
\label{introduction}

The cosmological significance of de Sitter space, both in its importance as a limit of the slow-roll regime of inflation and its possible role in the future evolution of the universe, has enhanced the need for a deeper understanding of interacting field theory in a de Sitter background.  The geometry of de Sitter space allows for a family of inequivalent vacuum states which can be parameterized by a complex number $\alpha$ and which are therefore called the $\alpha$-vacua \cite{tagirov1,tagirov2,mottola,allen}.  Among these vacua, only one---the Euclidean or Bunch-Davies vacuum \cite{bunch}---matches with the flat space vacuum at distances much shorter than the natural length scale associated with the curvature of de Sitter space.  Due to this property, this vacuum has been assumed to be the correct vacuum state during inflation.

The vacuum state during inflation is important since it can provide a window to Planck scale physics which is more sensitive to the new physics than effects coming from subleading operators calculated in the Euclidean vacuum.  From an effective theory perspective, higher dimension operators in the Euclidean vacuum \cite{kkls} generically give effects suppressed by $H^2/M^2$, where $M$ is the scale of new physics.  These would be unobservable unless the Hubble scale during inflation, $H$, is quite large.  However, depending on how the vacuum state is set at the scale $M$, the state during inflation is generally more sensitive to new physics; deviations from the cosmic microwave background (CMB) power spectrum for the Euclidean tree level result are only suppressed by $H/M$ \cite{ulf,kempf,gary,ss,bm,rich1,rich2}.  Loop effects in a non-thermal vacuum state can in some models be enhanced even further, scaling as $M/H$ \cite{loop}.  Once the CMB has been measured sufficiently well to see deviations from the tree-level Euclidean result, it will be important to separate what these deviations indicate about the inflationary Lagrangian and what they reveal about the vacuum state during inflation.

To address the role of non-thermal states in inflation, such as the $\alpha$-vacua of de Sitter space, requires a self-consistent construction for the quantum theory in one of these states.  For example, the usual description of perturbation theory in de Sitter space is guided by an intuition based on quantum field theory in Minkowski space.  The pathologies which appear for an interacting theory in an $\alpha$-vacuum, may indeed signal that the $\alpha$-vacuum states cannot be used other than as a free theory, but these pathologies might also indicate that the standard formulation of perturbation theory must be modified for an $\alpha$-vacuum.

The breakdown of an interacting scalar theory in an $\alpha$-vacuum is quite dramatic.  New divergences appear which, unlike those in the Euclidean vacuum, cannot be renormalized \cite{fate}.  Self-energy graphs in a theory with a cubic interaction produce pinched singularities \cite{einhorn1} or require peculiar non-local counterterms \cite{banks}.  To arrive at these pathologies, the Feynman propagator used for the original $\alpha$-vacuum is the Green's function associated with a single point source.  When the propagator is modified to include an additional source at the antipode as well, the self-energy diagrams are free of the non-renormalizable divergences \cite{lowe1,lowe2,einhorn2}.  The correlation of events at antipodal points renders the theory non-local. However, it is not immediately clear whether this non-locality leads to any inconsistencies; for example, whether observers in a restricted patch see any breakdown of causality or locality with respect to events within their horizon.

Here we study an interacting scalar field theory in a pure de Sitter background to learn whether the divergences that appear in the $\alpha$-vacuum can be tamed when the Feynman propagator or, equivalently, the generating functional of the theory is modified.  Since an interacting theory in a Euclidean vacuum is renormalizable, the only constraint is that these modifications should vanish in the Euclidean limit.  This construction generalizes and extends that which appears in \cite{lowe2,einhorn2}.  In this article, we shall derive the generating functional in more detail, demonstrating that theories with non-local interactions of the general form, 
\begin{eqnarray}
{\cal L} &=& 
{\textstyle{1\over 2}} g^{\mu\nu} \partial_\mu \Phi(x) \partial_\nu \Phi(x) 
- {\textstyle{1\over 2}} m^2 \Phi^2(x) 
\nonumber \\
&&
- \sum_{p+q\le 4} c_{p,q} \Phi^p(x) \Phi^q(x_A) , 
\label{pregenint}
\end{eqnarray}
where $x_A$ is the antipode to $x$, are causal and renormalizable at the level of the one loop self-energy diagrams.  Our construction adapts the Schwinger-Keldysh approach \cite{schwinger,keldysh,kt} to this setting, which is required since de Sitter space lacks an $S$-matrix description.

In the next section, we develop the path integral description for an interacting scalar field theory in a de Sitter background with sources at antipodal points.  This construction is first described in a coordinate-independent form.  Section \ref{inflpatch} derives the explicit form of the modified $\alpha$ propagator in the patch covered by inflationary coordinates.  In section \ref{renormalize}, we study the one loop self-energy correction from a cubic interaction for both a theory with local and a theory with a special set of antipodal interactions, showing that both are causal and renormalizable, before considering the most general set of antipodal interactions in section \ref{antipodes}.  Section \ref{conclude} presents our conclusions.

\section{A double-source $\alpha$-vacuum}
\label{doubled}

\subsection{The $\alpha$ propagator}

We shall explicitly refer to antipodal points when we construct the propagator for the $\alpha$-vacuum.  It is therefore initially convenient to use a coordinate system that covers the entire de Sitter space-time such as global coordinates \cite{houches}, 
\begin{equation}
ds^2 = d\tau^2 - \cosh^2\tau\, d\Omega^2 
\qquad \tau \in [-\infty, \infty] ,
\label{globalcoords}
\end{equation}
or conformal coordinates, 
\begin{equation}
ds^2 = {dT^2 - d\Omega^2\over\cos^2 T} 
\qquad T \in [-\pi/2, \pi/2] ;
\label{confcoords}
\end{equation}
$d\Omega^2$ is the metric of the 3-sphere.  If we denote the time coordinate generically by $t$, the antipodal map, $A: x \mapsto x_A$, is explicitly realized for either of these coordinate systems by 
\begin{equation}
A : (t,\Omega) \mapsto (-t, \Omega_A) , 
\label{antimap}
\end{equation}
where $\Omega_A$ represents the antipode on the 3-sphere to the point whose angular coordinates are represented by $\Omega$.

Consider a free massive scalar $\Phi(x)$ propagating in de Sitter space.  Expanding the field in a complete set of mode functions, $\phi_n^E(x)$,
\begin{equation}
\Phi(x) = \sum_n \left[ \phi_n^E(x)\, a_n^E + \phi_n^{E*}(x)\, a_n^{E\dagger} \right] , 
\label{genmodes}
\end{equation}
we can define the vacuum to be the state annihilated by $a_n^E$.  The mode functions are the solutions to the Klein-Gordon equation and the Euclidean vacuum state $|E\rangle$ is defined by the linear combination of the solutions to this equation that matches with the flat space vacuum modes when the curvature of de Sitter space is taken to vanish.  The Euclidean vacuum is also the unique vacuum whose Wightman function
\begin{equation}
G_E(x,x') = \langle E | \Phi(x)\Phi(x') | E \rangle 
= \sum_n \phi_n^E(x) \phi_n^{E*}(x') ,
\label{genWight}
\end{equation}
is analytic on the Euclidean sphere.  The Wightman function depends only on the de Sitter invariant distance between $x$ and $x'$ \cite{mottola,allen,bousso}.  To demonstrate later the de Sitter invariance of the $\alpha$-vacuum Wightman function, it is useful to choose the mode functions so that 
\begin{equation}
\phi_n^E(x_A) = \phi_n^{E*}(x) .
\label{antipodemodes}
\end{equation}

To study quantum field theory in this geometry, we introduce the Feynman propagator, $G_F^E(x,x')$.  Since the Euclidean vacuum becomes the flat space vacuum at short distances, it is most sensible to define the Feynman propagator as the Green's function associated with a single point source,
\begin{equation}
\bigl[ \nabla^2_x + m^2 \bigr] G_F^E(x,x') 
= {\delta^4(x-x')\over\sqrt{-g(x)}} ,  
\label{genEFeynKG}
\end{equation}
just as in flat space.  The solution to this equation is then given by 
\begin{eqnarray}
-i G_F^E(x,x') 
&\equiv& \langle E | T\bigl( \Phi(x)\Phi(x) \bigr) | E \rangle 
\nonumber \\
&=& \Theta(t-t')\, G_E(x,x') 
\nonumber \\
&&
+ \Theta(t'-t)\, G_E(x',x) .
\label{genEFeyn}
\end{eqnarray}

Despite being the unique state which is flat at short distances, the Euclidean vacuum is not the unique de Sitter invariant state.  Mottola \cite{mottola} and Allen \cite{allen} showed that there exists a one parameter family of vacua, $|\alpha\rangle$, defined by 
\begin{equation}
a_n^\alpha\, |\alpha\rangle = 0 , 
\label{alphaannih}
\end{equation}
where $a_n^\alpha$ is a Bogolubov transformation of the Euclidean operators, 
\begin{equation}
a_n^\alpha = N_\alpha \left[ a_n^E - e^{\alpha^*} a_n^{E\dagger} \right] . 
\label{alphadef}
\end{equation}
Here, $\alpha$ is a complex number with ${\rm Re}\, \alpha < 0$ and the normalization is chosen to preserve the form of the canonical commutation relation of the creation and annihilation operators of the $\alpha$-vacuum, 
\begin{equation}
N_\alpha = \left( 1 - e^{\alpha+\alpha^*} \right)^{-1/2} .
\label{normdef}
\end{equation}
The Wightman function associated with this vacuum, 
\begin{equation}
G_\alpha(x,x') = \langle\alpha | \Phi(x)\Phi(x') | \alpha\rangle , 
\label{genAWight}
\end{equation}
also only depends on the de Sitter invariant distance $Z(x,x')$ between $x$ and $x'$.  The reason is that the distance between $x_A$ and $x'$ can be expressed in terms of the distance between $x$ and $x'$,
\begin{equation}
Z(x_A,x') = - Z(x,x') . 
\label{dSdist}
\end{equation}
The transformation that defines $a_n^\alpha$ also relates the mode functions for the $\alpha$-vacuum to those of the Euclidean vacuum, 
\begin{equation}
\phi_n^\alpha(x) = N_\alpha \left[ \phi_n^E(x) + e^\alpha \phi_n^{E*}(x) \right] .
\label{phialphadef}
\end{equation}
Using this equation with Eq.~(\ref{antipodemodes}), the $\alpha$ Wightman function becomes 
\begin{eqnarray}
G_\alpha(x,x') 
&=& N_\alpha^2 \Bigl[ G_E(x,x') + e^{\alpha+\alpha^*} G_E(x',x) 
\nonumber \\
&&
+ e^\alpha G_E(x_A,x') + e^{\alpha^*} G_E(x',x_A) \Bigr] . \qquad
\label{genAWightasE}
\end{eqnarray}
In this form, the de Sitter invariance of the Euclidean Wightman function combined with Eq.~(\ref{dSdist}) implies that the $\alpha$ Wightman function also only depends on $Z(x,x')$.

The proper definition of the Feynman propagator in an $\alpha$-vacuum is more subtle.  If we were to define the time-ordered propagator in the $\alpha$-vacuum by a straightforward generalization of the Euclidean vacuum definition, i.e. 
\begin{eqnarray}
\langle\alpha | T \bigl( \Phi(x)\Phi(x') \bigr) | \alpha\rangle
&{\buildrel ? \over \sim}& 
\Theta(t-t')\, G_\alpha(x,x') 
\nonumber \\
&&
+ \Theta(t'-t)\, G_\alpha(x',x) , \qquad
\label{badAdef}
\end{eqnarray}
then the appearance of the opposite ordering of the points in the pairs of terms in the first and second lines of Eq.~(\ref{genAWightasE}) would lead to incompatible orderings in Eq.~(\ref{badAdef}).  The effects of this pathology become apparent when we attempt to use this propagator as a basis for perturbation theory in the $\alpha$-vacuum.  When studying the loop corrections in an interacting theory, interference between the opposite orderings produce either pinched singularities \cite{einhorn1} or non-renormalizable divergences \cite{fate}.

From a global perspective, the $\alpha$ Wightman function contains antipodal information but the time-ordering, as indicated by the $\Theta$-functions in Eq.~(\ref{badAdef}), does not.  This inconsistency suggests that we should generalize the time-ordering used for propagation in an $\alpha$-vacuum so that we obtain a renormalizable theory.  

We construct a new $\alpha$-dependent operator $T_\alpha$ to disentangle the conflicting time-orderings which would occur in Eq.~(\ref{badAdef}).  Since the Euclidean vacuum smoothly matches onto the Minkowski vacuum, $T_\alpha$ should revert to the usual definition in the Euclidean limit,
\begin{equation}
\lim_{\alpha\to -\infty} \langle\alpha | T_\alpha\left( \Phi(x)\Phi(x') \right) | \alpha \rangle = -i G_F^E(x,x') . 
\label{limitgreens}
\end{equation}
The constructive interference of the opposite point-orderings in Eq.~(\ref{genAWightasE}) is avoided by defining
\begin{eqnarray}
T_{\alpha} \bigl( \Phi(x) \Phi(x') \bigr)
&=& \Theta_\alpha(t,t')\, \Phi(x) \Phi(x') 
\nonumber \\
&&
+ [\Theta_\alpha(t',t)]^*\, \Phi(x') \Phi(x)
\nonumber \\
&&
+\, \Theta^A_\alpha(t_A,t')\, \Phi(x_A) \Phi(x') 
\nonumber \\
&&
+ [\Theta_\alpha^A(t',t_A)]^*\, \Phi(x') \Phi(x_A) 
\qquad
\label{Talphadef}
\end{eqnarray}
where 
\begin{eqnarray}
\Theta_\alpha(t,t') 
&\equiv& {1\over 1-e^{2\alpha}} \Bigl[ 
A_\alpha 
\left[ \Theta(t-t') + e^{2\alpha} \Theta(t'-t) \right]
\nonumber \\
&&\qquad
- B_\alpha e^\alpha 
\left[ \Theta(t_A-t') + \Theta(t'-t_A) \right] \Bigr] 
\nonumber \\
\Theta_\alpha^A(t_A,t') 
&\equiv& {1\over 1-e^{2\alpha}} \Bigl[ 
B_\alpha
\left[ \Theta(t_A-t') + e^{2\alpha} \Theta(t'-t_A) \right] 
\nonumber \\
&&\qquad
- A_\alpha e^\alpha 
\left[ \Theta(t-t') + \Theta(t'-t) \right] \Bigr] . 
\nonumber \\
\label{ThetaAdefs}
\end{eqnarray}
The weighting factors, $A_\alpha$ and $B_\alpha$, are assumed to be real.  As long as 
\begin{equation}
\lim_{\alpha\to -\infty} A_\alpha = 1 
\quad\hbox{and}\quad
\lim_{\alpha\to -\infty} B_\alpha = 0 , 
\label{Qalimits}
\end{equation}
 $T_\alpha$ reduces to the ordinary time-ordering operator.  This choice is the essentially unique time ordering prescription---up to the choice of the constants $A_\alpha$ and $B_\alpha$---that does not lead to a phase-cancelling interference in products of propagators.  

The expectation value of the $T_\alpha$-ordered product of two fields defines the propagator for the $\alpha$-vacuum.  Despite the complicated form of Eq.~(\ref{ThetaAdefs}), the $\alpha$ propagator has a simple expression in terms of the Euclidean propagators; using Eq.~(\ref{Talphadef}) and Eq.~(\ref{genAWightasE}) yields 
\begin{eqnarray}
&&\langle\alpha | T_{\alpha} \bigl( \Phi(x) \Phi(x') \bigr) | \alpha\rangle
\nonumber \\
&&\qquad
= -i A_\alpha G_E^F(x,x') - i B_\alpha G_E^F(x_A,x') . \quad
\label{alphaprop}
\end{eqnarray}
If we define the $\alpha$ Feynman propagator by 
\begin{equation}
\langle\alpha | T_{\alpha} \bigl( \Phi(x) \Phi(x') \bigr) | \alpha\rangle
\equiv -i G_\alpha^F(x,x') , 
\label{afeyndef}
\end{equation}
then from the equation for the Euclidean propagator (\ref{genEFeynKG}) we see that the $\alpha$ propagator corresponds to having {\it two\/} sources, at $x$ and its antipode $x_A$, weighted appropriately,
\begin{equation}
\bigl[ \nabla^2_x + m^2 \bigr] G_\alpha^F(x,x') 
= A_\alpha {\delta^4(x-x')\over\sqrt{-g(x)}} 
+ B_\alpha {\delta^4(x_A-x')\over\sqrt{-g(x_A)}} . 
\label{afeynKG}
\end{equation}

Notice that when 
\begin{eqnarray}
A_\alpha &=& N_\alpha^2 (1 + e^{\alpha+\alpha^*}) 
\nonumber \\
B_\alpha &=& N_\alpha^2 (e^\alpha + e^{\alpha^*}) 
\label{GLdef}
\end{eqnarray}
we have the propagator structure obtained in \cite{lowe2} using a real time-ordering prescription while 
\begin{eqnarray}
A_\alpha &=& N_\alpha^2 (1 + e^{\alpha+\alpha^*}) 
\nonumber \\
B_\alpha &=& 2 N_\alpha^2 e^\alpha 
\label{ELdef}
\end{eqnarray}
corresponds to the propagator for the `squeezed states' of \cite{einhorn2}.  The propagator defined by Eq.~(\ref{afeynKG}) contains sources at both $x$ and its antipode $x_A$ unlike the original definition of the Feynman propagator in \cite{mottola,allen,einhorn1,banks,fate} which only included a source at $x$.  Although this correlation of events at a point and its antipode is manifestly non-local, this non-locality is of a restricted, global form since a generic point in de Sitter space and its antipode are always causally disconnected.  

The action of $T_\alpha$ on products of more than two fields can be defined implicitly once we have constructed a path integral formulation which reproduces Eq.~(\ref{afeyndef}).  One of the advantages of defining $T_\alpha$ as above is that it acts on the fields directly, before we have taken the expectation value in an $\alpha$-vacuum, unlike the time-orderings \cite{einhorn2} which were defined to act on matrix elements.

\subsection{The path integral:  free theory}

The double source propagator for the $\alpha$-vacuum has a natural formulation in terms of a path integral.  The construction of the path integral is useful since once we have found the proper form for the free field generating functional, it is a simple matter to generalize to the interacting case for the Schwinger-Keldysh formalism.  In this section we find the generating functional which yields the $\alpha$ propagator introduced in the Eq.~(\ref{afeyndef}).

Since the $\alpha$ propagator contains two sources, let us define the generating functional to be 
\begin{equation}
W_0^\alpha[J] = \int {\cal D}\Phi\, e^{i \int d^4x\, \sqrt{-g}\, 
\left[ {\cal L}_0 + \left( a_\alpha J(x) + b_\alpha J(x_A) \right) \Phi(x) \right] } . 
\label{genfunc}
\end{equation}
The free Lagrangian is 
\begin{equation}
{\cal L}_0 = {\textstyle{1\over 2}} g^{\mu\nu}\partial_\mu\Phi \partial_\nu\Phi - {\textstyle{1\over 2}} m^2 \Phi^2 .
\label{freeLagrange}
\end{equation}
The current couples to the field at both $x$ and $x_A$ with weightings $a_\alpha$ and $b_\alpha$ respectively.  Matching onto the Euclidean limit fixes $a_\alpha\to 1$ and $b_\alpha\to 0$ as $\alpha\to -\infty$, so that we obtain the correct Euclidean propagator,
\begin{eqnarray}
&&\!\!\!\!\!\!\!\!\!\!\!\!\!\!\!\!\!\!\!\!\!\!\!\!
\langle E | T \bigl( \Phi(x)\Phi(x') \bigr) | E \rangle 
\nonumber \\
&=& \left[ -i {\textstyle{\delta\over\delta J(x)}} \right] 
\left[ -i {\textstyle{\delta\over\delta J(x')}} \right] 
\left. W_0^E[J] \right|_{J=0} 
\nonumber \\
&=& 
\int {\cal D}\Phi\, \Phi(x)\Phi(x') e^{i \int d^4y\, \sqrt{-g}\, {\cal L}_0 } . 
\label{PIEprop}
\end{eqnarray}
The $\alpha$ propagator is then obtained by taking the functional derivative of the two source generating functional,
\begin{eqnarray}
&&\!\!\!\!\!\!\!\!\!\!\!\!\!\!\!\!\!\!\!\!\!\!\!\!
\left[ -i {\textstyle{\delta\over\delta J(x)}} \right] 
\left[ -i {\textstyle{\delta\over\delta J(x')}} \right] \left. W_0^\alpha[J] \right|_{J=0} 
\nonumber \\
&=& 
(a_\alpha^2+b_\alpha^2)\, \langle E | T \bigl( \Phi(x)\Phi(x') \bigr) | E \rangle 
\nonumber \\
&&
+ 2a_\alpha b_\alpha\, \langle E | T \bigl( \Phi(x_A)\Phi(x') \bigr) | E \rangle . 
\label{funcderiv}
\end{eqnarray}
Here we have used the relation
\begin{equation}
\langle E | T \bigl( \Phi(x_A)\Phi(x'_A) \bigr) | E \rangle 
= \langle E | T \bigl( \Phi(x)\Phi(x') \bigr) | E \rangle , 
\label{timeordants} \\
\end{equation}
assuming that the ordering of the antipodal times is the opposite of that of the original points.  Matching Eq.~(\ref{funcderiv}) with Eq.~(\ref{alphaprop}) determines the weightings for the currents,
\begin{eqnarray}
a_\alpha &=& {1\over\sqrt{2}} \left[ A_\alpha 
+ \sqrt{ A_\alpha^2 - B_\alpha^2 }
\right]^{1/2}
\nonumber \\
b_\alpha &=& {1\over 2} {B_\alpha\over a_\alpha} , 
\label{qsasQs}
\end{eqnarray}
which fixes the path integral definition of the propagator in the $\alpha$-vacuum,
\begin{eqnarray}
&&\!\!\!\!\!\!\!\!\!\!\!\!\!\!\!\!\!\!\!\!\!\!\!\!
\langle\alpha | T_\alpha \bigl( \Phi(x)\Phi(x') \bigr) | \alpha\rangle
\nonumber \\
&=& \left[ -i {\textstyle{\delta\over\delta J(x)}} \right] 
\left[ -i {\textstyle{\delta\over\delta J(x')}} \right] 
\left. W_0^\alpha[J] \right|_{J=0} . 
\label{pathintdef}  
\end{eqnarray}

The Wightman functions for the theory are still essentially those of the usual $\alpha$-vacuum.  What has been altered is the definition of the propagator---from the perspective of the path integral formulation---or the time-ordering---from the canonical perspective.  Since the propagator can be written in terms of the Euclidean propagators, $G_E^F(x,x')$ and $G_E^F(x_A,x')$, the $\alpha$ propagator is still manifestly de Sitter invariant.  

The generating functional can alternately be written without the coupling between the field and source by completing the square in the exponent,
\begin{eqnarray}
\Phi'(x) &=& \Phi(x) - {1\over a_\alpha^2 - b_\alpha^2} 
\int d^4y\, \sqrt{-g}\, G_\alpha^F(x,y) \quad
\label{fieldshift} \\
&&\qquad\qquad\qquad\qquad
\times\left[ a_\alpha\, J(y) - b_\alpha\, J(y_A) \right],
\nonumber 
\end{eqnarray}
so that
\begin{equation}
W_0^\alpha[J] = {\cal N}_0 e^{{i\over 2}\int d^4x\, \sqrt{-g}\int d^4y\, \sqrt{-g}\,  J(x) G_\alpha^F(x,y) J(y) } 
\label{genfuncJJ}
\end{equation}
with 
\begin{equation}
{\cal N}_0 \equiv \int {\cal D}\Phi'\, 
e^{-{i\over 2}\int d^4x\, \sqrt{-g}\,  \Phi' [\nabla^2 + m^2] \Phi'} . 
\label{gennorm}
\end{equation}
To obtain the correctly normalized propagator starting from either form of the generating functional, Eq.~(\ref{genfunc}) or Eq.~(\ref{genfuncJJ}), set ${\cal N}_0=1$.

\subsection{The path integral:  interactions}

In an interacting theory, the generating functional is given by 
\begin{equation}
W^\alpha[J] = {\cal N} \int {\cal D}\Phi\, e^{i \int d^4x\, \sqrt{-g}\, 
\left[ {\cal L} + \left( a_\alpha J(x) + b_\alpha J(x_A) \right) \Phi(x) \right] } . 
\label{fgenfunc}
\end{equation}
with
\begin{equation}
{1\over {\cal N}} =  \int {\cal D}\Phi\, e^{i \int d^4x\, \sqrt{-g}\, 
{\cal L} } 
\label{genvactovac}
\end{equation}
where now the Lagrangian contains an interacting piece,
\begin{equation}
{\cal L}[\Phi(x)] = {\cal L}_0[\Phi(x)] + {\cal L}_I[\Phi(x)] . 
\label{intLagrang}
\end{equation}
Note that here we have assumed that ${\cal L}_I$ only includes local interactions.  We next define the following functional derivative
\begin{equation}
{\delta\over\delta{\cal J}(x)} \equiv {1\over a_\alpha^2 - b_\alpha^2} 
\left[ a_\alpha {\delta\over\delta J(x)} - b_\alpha {\delta\over\delta J(x_A)} \right] , 
\label{ddtJ}
\end{equation}
which allows us to separate the interacting piece of the Lagrangian from the free theory generating functional,
\begin{equation}
W^\alpha[J] = {\cal N} 
e^{i \int d^4x\, \sqrt{-g}\, {\cal L}_I 
\left[ -i {\delta\over\delta{\cal J}(x)}\right] } W_0^\alpha[J] . 
\label{fgenfuncJt}
\end{equation}

Because a different functional derivative is used to produce local interactions and to extract a field, $T_\alpha$-ordered matrix elements involve several types of propagators depending on whether it is external or wholly internal.  For example, a general $n$-point Green's function is given by 
\begin{eqnarray}
G_\alpha^n(x_1,\ldots , x_n) 
&=&
\langle\alpha |\, T_\alpha \bigl( \Phi(x_1) \cdots \Phi(x_n) \bigr)\, |\alpha\rangle 
\label{greens} \\
&=&
\left[ -i {\textstyle{\delta\over\delta J(x_1)}} \right]
\cdots 
\left[ -i {\textstyle{\delta\over\delta J(x_n)}} \right] 
\nonumber \\
&&
{\cal N} 
e^{i \int d^4x\, \sqrt{-g}\, {\cal L}_I 
\left[ -i {\delta\over\delta{\cal J}(x)}\right] } W_0^\alpha[J] 
\Bigr|_{J=0} . 
\nonumber 
\end{eqnarray}
The propagator between two internal points, because it only contains ${\cal J}$-derivatives, 
\begin{equation}
\left[ -i {\textstyle{\delta\over\delta{\cal J}(x)}} \right] 
\left[ -i {\textstyle{\delta\over\delta{\cal J}(x')}} \right] 
W_0^\alpha[J] \Bigr|_{J=0} 
= -i G_E^F(x,x') , 
\label{JtJtG}
\end{equation}
actually yields a Euclidean propagator while a propagator between an internal and an external point produces a mixed propagator,
\begin{eqnarray}
-i G_m^F(x,x') &\equiv&
\left[ -i {\textstyle{\delta\over\delta J(x)}} \right] 
\left[ -i {\textstyle{\delta\over\delta{\cal J}(x')}} \right] 
W_0^\alpha[J] \Bigr|_{J=0} 
\nonumber \\
&=& -i \left[ a_\alpha G_E^F(x,x') + b_\alpha G_E^F(x_A,x') \right] . 
\qquad\ 
\label{JJtG}
\end{eqnarray}
It is the fact that the internal propagators are Euclidean that leads to the renormalizability of the Green's functions based on the double source generating functional of Eq.~(\ref{fgenfunc}).  All the $\alpha$-dependence in the Green's functions is generated by the external legs.

In writing the interacting generating functional in Eq.~(\ref{fgenfuncJt}), we have assumed that all the terms in the interacting part of the Lagrangian are local.  However, in order to produce a propagator that does not produce the the divergences seen in the original $\alpha$-vacuum of \cite{mottola,allen}, we have already included a specific non-local source in the term $J(x_A)\Phi(x)$.  If we relax our requirement on the locality of the interactions as well, we can include interactions of the form, 
\begin{equation}
{\cal L}_I = \sum_n c_n \widetilde\Phi^n(x) .
\label{specialint}
\end{equation}
where we have defined
\begin{equation}
\widetilde\Phi(x) \equiv a_\alpha\Phi(x) + b_\alpha\Phi(x_A) .
\label{specialfield}
\end{equation}
Note that the coefficients of the two terms on the right side of this definition are the same as the coefficients of the non-local source terms in the free field generating functional in Eq.~(\ref{genfunc}).  Such theories are particularly interesting \cite{lowe2} since then the generating functional has a simple structure 
\begin{equation}
W^\alpha_A[J] = {\cal N} 
e^{i \int d^4x\, \sqrt{-g}\, \sum_n c_n 
\left[ -i {\delta\over\delta J(x)}\right]^n } W_0^\alpha[J] . 
\label{fgenfuncnoJt}
\end{equation}
In this case, both internal and external propagators are the $\alpha$ propagators given in Eq.~(\ref{alphaprop}).  Here, the renormalizability of the theory is less trivial since internal loops are not only composed of Euclidean propagators as in Eq.~(\ref{fgenfuncJt}).  Later we shall examine the Green's functions for a theory with these special non-local interactions to show that they too are causal and renormalizable before we consider the most general interacting theory containing arbitrary products of powers of $\Phi(x)$ and $\widetilde\Phi(x)$.

\subsection{Finite time evolution} 

So far we have written the generating functional appropriate for an $S$-matrix calculation, which is ill-defined for de Sitter space.  The Schwinger-Keldysh \cite{schwinger,keldysh,kt} formalism evades the necessity of using asymptotic `in' and `out' states by solving for the evolution of a matrix element from a given initial state defined at an arbitrary time.  The application of this formalism to de Sitter space is described in detail in \cite{fate}.  

\begin{figure}[!tbp]
\includegraphics{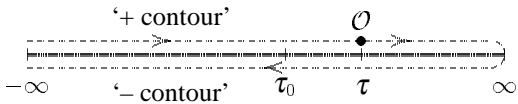}
\caption{The contour used to evaluate the evolution of operators over a finite time interval.  Here we have shown the contour doubling for global coordinates, Eq.~(\ref{globalcoords}), but an analogous prescription is used for any coordinate system.  The initial state is specified at $\tau_0$ and is evolved until $\tau$.  We double the field content so that separate copies of the fields are used for the upper and lower parts of the contour.\label{ctp}}
\end{figure}

The principal difference between the Schwinger-Keldysh and the $S$-matrix approaches is that in the latter we evaluate the expectation values between an `in' state and an `out' state, which is obtained by acting upon the `in' state with a single time-evolution operator, $U(\infty,-\infty)$,
\begin{equation}
\langle {\rm out} | {\cal O}(t) | {\rm in}\rangle 
= \langle {\rm in} | U^\dagger(\infty,-\infty){\cal O}(t) | {\rm in}\rangle . 
\label{inout}
\end{equation}
In flat space this is a sensible definition since we have a global time-like Killing vector through which we can relate these asymptotic `in' and `out' states.  In the Schwinger-Keldysh formulation, we instead evaluate the expectation value of an operator at a time $t$ by time evolving both states so that we are essentially examining an `in'-`in' expectation value, 
\begin{eqnarray}
&&\!\!\!\!\!\!\!\!\!\!\!\!\!\!\!\!\!\!\!\!\!\!\!\!
\langle \Psi'(t) | {\cal O}(t) | \Psi(t) \rangle 
\nonumber \\
&=& \langle \Psi'(t_0) | U^\dagger(\infty,t_0){\cal O}(t) U(\infty,t_0) | \Psi(t_0) \rangle . 
\label{outout}
\end{eqnarray}
Since both states are time-evolved from initial states specified at $t_0$, the matrix element includes two insertions of the time evolution operator, one of which is conjugated since it acts on the $\langle\Psi'|$ state.  To simplify the calculation, both time-evolution operators can be formally combined into a single time-ordered operator, which we integrate over a closed time contour from $t_0$ to $\infty$ and then back to $t_0$, as shown in Fig.~\ref{ctp}.  
This procedure effectively doubles the field content of the theory with the fields associated with the forward part of the contour labeled by $\Phi^+$ while fields on the reversed part of the contour are labeled by $\Phi^-$.  Since the $\infty$ to $t_0$ part of the contour on which the $\Phi^-$ fields live is associated with the $\langle\Psi'|$ state, events on this part of the contour are assumed to always occur after and in the opposite order as those on the $t_0$ to $\infty$ portion of the contour.  The reversal of the time ordering follows from the conjugation of the operator $U^\dagger(\infty,t_0)$.  Note that the two time evolution operators associated with the $\langle\Psi'|$ state and the $|\Psi\rangle$ state cancel over the interval from $t$ to $\infty$ so that the evolution remains causal.

To incorporate this approach for evaluating matrix elements into a path integral representation, we generalize the generating functional to include currents for both parts of the contour, $J^+(x)$ and $J^-(x)$.  The Schwinger-Keldysh generating functional is then given by 
\begin{eqnarray}
\ &&\!\!\!\!\!\!\!\!\!\!\!\!\!\!\!\!\!\!\!\!\!\!\!\!
W^\alpha[J^+,J^-] 
\label{SKgenfunc} \\
&=& {\cal N} 
e^{i \int_{t_0}^{t'} d^4x\, \sqrt{-g}\, \left\{ 
{\cal L}_I \left[ -i {\delta\over\delta{\cal J}^+(x)}\right] 
- {\cal L}_I \left[ -i {\delta\over\delta{\cal J}^-(x)}\right] \right\} } 
\nonumber \\
&&\quad
\times W_0^\alpha[J^+,J^-] . 
\nonumber
\end{eqnarray}
Here the most useful expression for the free generating functional is given by generalizing Eq.~(\ref{genfuncJJ}), 
\begin{equation}
W_0^\alpha[J^+,J^-] = e^{{i\over 2}\int d^4x\, \sqrt{-g}\int d^4x'\, \sqrt{-g}\,  J^a(x) G_\alpha^{ab}(x,x') J^b(x') } 
\label{genfuncJJpm}
\end{equation}
where the labels refer to the upper or lower part of the contour, $a,b = +,-$.  The time-ordering along the contours is assumed to be preserved under the antipode mapping in a global coordinate system.

The case of a coordinate patch which only covers part of de Sitter space is more subtle since the antipodal time may lie in a different coordinate patch.  However, such cases must be addressed since it is possible to define an $\alpha$-vacuum even in a non-global coordinate system and moreover global coordinates are rarely used for cosmological calculations.  In the next section, we construct the correct time orderings needed for conformally flat coordinates which cover the same region of de Sitter space as inflationary coordinates.  There, we define the proper form of the contour propagators, $G_\alpha^{ab}(x,x')$.

\section{The inflationary patch}
\label{inflpatch}

The standard coordinates used in inflation, 
\begin{equation}
ds^2 = dt^2 - e^{2Ht}\, d\vec x^2  
\qquad
t\in [-\infty,\infty] ,
\label{inflatemetric}
\end{equation}
can be written in a conformally flat form, 
\begin{equation}
ds^2 = {d\eta^2 - d\vec x^2\over H^2\eta^2} 
\qquad
\eta\in [-\infty,0] ,
\label{metric}
\end{equation}
by defining $\eta = - H^{-1}e^{-Ht}$.  $H$ is the Hubble constant and is related to the cosmological constant by $\Lambda = 6H^2$.  This coordinate system covers only half of de Sitter space.  The other half of the space is covered by a set of coordinates with $\eta\to -\eta$.  

\begin{figure}[!tbp]
\includegraphics{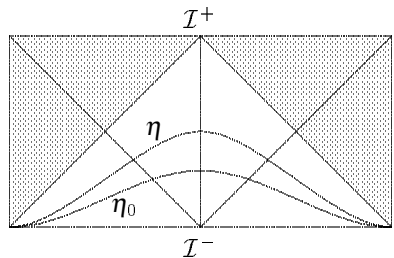}
\caption{The Penrose diagram for de Sitter space showing the region covered by inflationary coordinates which has been left unshaded.  The past and future null infinity surfaces are denoted by ${\cal I}^-$ and ${\cal I}^+$.  The Schwinger-Keldysh approach evolves a matrix element for an initially specified state at $\eta_0$ forward to an arbitrary later time $\eta$.
\label{evolve}}
\end{figure}

In principle, we should use separate sets of spatial coordinates on each patch,
\begin{equation}
x\in {\rm dS}_4 = \cases{ 
(\eta, \vec x) &if $\eta\in [-\infty, 0]$ \cr
(\eta, \vec x_{\scriptscriptstyle A}) &if $\eta\in [0, \infty ]$ \cr} . 
\label{northsouth}
\end{equation}
Here the subscript on $\vec x_{\scriptscriptstyle A}$ is only a label and does not correspond to taking any spatial antipode as was done in the case of the global coordinates, Eq.~(\ref{antimap}).  Under the antipodal map 
\begin{equation}
A:(\eta, \vec x) \mapsto 
(-\eta, \vec x_{\scriptscriptstyle A}) . 
\label{Anorthsouth}
\end{equation}
We arrange our coordinatization of the $\eta\in[0,\infty]$ patch so that the antipode is labeled formally by the same coordinate value in the spatial part, $\vec x_{\scriptscriptstyle A} = \vec x$.  In studying quantum field theory in a de Sitter background, we shall only consider events $x$ in the patch with $\eta\in[-\infty,0]$ where the only information from the other patch appears through the image sources at $x_A$ used to construct the $\alpha$ propagator.  The generating functional can then be written in terms of conformal time integrals only over half of de Sitter space, 
\begin{eqnarray}
W_0^\alpha[J^+,J^-] &=&
\exp\biggl\{ {i\over 2}
\int_{\eta_0}^0 {d\eta\over\eta^4} 
\int_{\eta_0}^0 {d\eta'\over\eta^{\prime 4}} 
\int d^3\vec x d^3\vec x'
\label{genfuncJJ4} \\
&&\quad
\bigl[ 
J^a(\eta,\vec x) G_\alpha^{ab}(\eta,\vec x;\eta',\vec x') J^b(\eta',\vec x') 
\bigr]
\biggr\} . 
\nonumber
\end{eqnarray}
The interacting theory generating functional based on $W_0^\alpha[J^+,J^-]$ evolves a state defined at an initial time surface, $\eta_0$, to some future time $\eta$ as shown in Fig.~\ref{evolve}.  This approach avoids the need to define asymptotic `in' and `out' states on the past and future null infinity surfaces ${\cal I}^-$ and ${\cal I}^+$, respectively.

In conformally flat coordinates the spatial symmetry greatly simplifies the form of the mode expansion of a free scalar field,
\begin{equation}
\Phi(\eta,\vec x) = \int {d^3\vec k\over(2\pi)^3}\, 
\left[ U_k^E(\eta) e^{i\vec k\cdot \vec x} a_{\vec k}^E 
+ U_k^{E*}(\eta) e^{-i\vec k\cdot \vec x} a_{\vec k}^{E\dagger} 
\right] , 
\label{Eexpand}
\end{equation}
where the mode functions $U_k^E(\eta)$ satisfy the Klein-Gordon equation,
\begin{equation}
\left[ \eta^2 \partial_\eta^2 - 2 \eta \partial_\eta + \eta^2 k^2 + m^2 H^{-2} \right] U_k^E(\eta) = 0 , 
\label{KGeqn}
\end{equation}
and where $k=|\vec k|$.  The correct linear combination of the solutions to this equation that matches onto the Minkowski space vacuum in the $H\to 0$ limit is, up to an arbitrary phase, \cite{fate}
\begin{equation}
U_k^E(\eta) = {\sqrt{\pi}\over 2} H \eta^{3/2} H_\nu^{(2)}(k\eta) 
\label{Emodes}
\end{equation}
where $H_\nu^{(2)}(k\eta)$ is a Hankel function with 
\begin{equation}
\nu = \sqrt{{9\over 4} - {m^2\over H^2}} . 
\label{nudef}
\end{equation}
Once we have determined the form of the Euclidean mode functions, we set $H=1$.  The spatial flatness of the metric also ensures that the Wightman function only depends on the difference of the spatial coordinates; in the momentum representation of the spatial part, the Wightman function becomes
\begin{equation}
G_E(x,x') = \int {d^3\vec k\over (2\pi)^3}\, e^{i\vec k\cdot (\vec x - \vec x')} G_k^E(\eta,\eta') 
\label{FTwight}
\end{equation}
with 
\begin{equation}
G_k^E(\eta,\eta') = U_k^E(\eta) U_k^{E*}(\eta')
\label{FTwightman}
\end{equation}
The momentum representation of the Euclidean Feynman propagator in conformally flat coordinates, 
\begin{equation}
G_k^F(\eta,\eta') = \int d^3\vec x\, e^{-i\vec k\cdot (\vec x-\vec x')} G_E^F(\eta,\vec x; \eta',\vec x') , 
\label{FTfeyn}
\end{equation}
which solves Eq.~(\ref{genEFeynKG}), is 
\begin{equation}
G_k^F(\eta,\eta') = \Theta(\eta-\eta')\, G_k^>(\eta,\eta') + \Theta(\eta'-\eta)\, G_k^<(\eta,\eta') . 
\label{FTfeynk}
\end{equation}
From the Schwinger-Keldysh perspective, it is useful to denote the different time-orderings in the Wightman functions with the $^>$($^<$) superscript rather than in the arguments,
\begin{eqnarray}
G_k^>(\eta,\eta') &=& i G_k^E(\eta,\eta') = iU_k^E(\eta)U_k^{E*}(\eta')
\nonumber \\
G_k^<(\eta,\eta') &=& i G_k^E(\eta',\eta) = iU_k^E(\eta')U_k^{E*}(\eta) . 
\label{Ggtltdef}
\end{eqnarray}

In addition to a Feynman propagator between points in the same coordinate patch, the $\alpha$-vacuum propagator in Eq.~(\ref{alphaprop}) also contains a propagator between $x_A$ and $x'$.  To represent these terms, we can use the same mode functions as above evaluated at the `antipodal time', 
\begin{equation}
G_E^F(x_A, x') = \int {d^3\vec k\over (2\pi)^3}\, e^{i\vec k\cdot (\vec x-\vec x')} G_k^F(-\eta,\eta') , 
\label{FTfeynA}
\end{equation}
where
\begin{eqnarray}
G_k^F(-\eta,\eta') &=& \Theta(-\eta-\eta')\, G_k^>(-\eta,\eta') 
\nonumber \\
&&
+ \Theta(\eta'+\eta)\, G_k^<(-\eta,\eta') . 
\label{FTfeynkApre}
\end{eqnarray}
In the conformally flat patch, $\eta,\eta'<0$ so that the first $\Theta$-function is always unity while the second always vanishes.  Thus
\begin{equation}
G_k^F(-\eta,\eta') = G_k^>(-\eta,\eta') = iU_k^E(-\eta)U_k^{E*}(\eta') . 
\label{FTfeynkA}
\end{equation}
To obtain mode functions which behave correctly under the antipodal map defined in Eq.~(\ref{Anorthsouth}), note that the mode functions $\tilde U_k^E(\eta_A)$ for the antipodal patch, where $\eta_A\in [0,\infty]$, satisfy the same Klein-Gordon equation (\ref{KGeqn}).  From the reversal of the sign of the time coordinate in the antipodal patch, the correct linear combination of the general solutions to the Klein-Gordon equation which matches onto the flat vacuum at short distances is
\begin{equation}
\tilde U_k^E(\eta_A) = e^{i\pi\left( \nu - {1\over 2} \right)} 
{\sqrt{\pi}\over 2} \eta_A^{3/2} H_\nu^{(1)}(k\eta_A) . 
\label{EmodesA}
\end{equation}
The phase has been chosen so that these mode functions match those in the $\eta\in[-\infty,0]$ patch under the antipodal map, $\eta_A=-\eta$,
\begin{equation}
U_k^E(-\eta) = \tilde U_k^E(-\eta) , 
\label{EEAmodes}
\end{equation}
so that we can formally extend the mode functions $U_k^E(\eta)$ to encompass the entire de Sitter space-time.

The important feature of the point-antipode part of the $\alpha$ propagator is that its behavior is fixed {\it before\/} extending to the full closed time path.  In the specific example discussed in the next section, we shall see that this choice establishes the correct convention to obtain a causal evolution of the matrix elements over finite intervals.  Therefore, the four contour propagators which appear in the free generating functional in Eq.~(\ref{genfuncJJ4}) are  
\begin{eqnarray}
G_k^{++}(\eta,\eta') 
&=& {\cal A} \bigl[ \Theta(\eta-\eta')\, G_k^>(\eta,\eta') 
\nonumber \\
&&\quad
+ \Theta(\eta'-\eta)\, G_k^<(\eta,\eta') \bigr]
+ {\cal B}\, G_k^>(-\eta,\eta') 
\nonumber \\
G_k^{--}(\eta,\eta') 
&=& {\cal A} \bigl[ \Theta(\eta'-\eta)\, G_k^>(\eta,\eta') 
\nonumber \\
&&\quad
+ \Theta(\eta-\eta')\, G_k^<(\eta,\eta') \bigr]
+ {\cal B}\, G_k^>(-\eta,\eta') 
\nonumber \\
G_k^{-+}(\eta,\eta') 
&=& {\cal A}\, G_k^>(\eta,\eta') + {\cal B}\, G_k^>(-\eta,\eta')
\nonumber \\
G_k^{+-}(\eta,\eta') 
&=& {\cal A}\, G_k^<(\eta,\eta') + {\cal B}\, G_k^>(-\eta,\eta') \quad
\label{countourGreens}
\end{eqnarray}
where 
\begin{equation}
{\cal A} = 1 \qquad {\cal B} = 0 
\label{Ecase}
\end{equation}
for a Euclidean propagator,
\begin{equation}
{\cal A} = a_\alpha \qquad {\cal B} = b_\alpha 
\label{Mcase}
\end{equation}
for a mixed propagator and
\begin{equation}
{\cal A} = A_\alpha \qquad {\cal B} = B_\alpha 
\label{Acase}
\end{equation}
for an $\alpha$ propagator.

It is helpful in some instances to consider a massless, conformally coupled scalar field, which corresponds to an effective mass parameter of $m^2=2$ or $\nu={1\over 2}$.  In this case, the Euclidean mode functions simplify to 
\begin{equation}
U_k^E(\eta)\bigr|_{\nu=1/2} = {i\over\sqrt{2k}} \eta e^{-ik\eta} . 
\label{UEhalfnu}
\end{equation}
In terms of the de Sitter invariant distance, 
\begin{equation}
Z(x,x') = {\eta^2 + \eta^{\prime 2} - |\vec x-\vec x'|^2 \over 2\eta\eta'} , 
\label{dSinvinCflat}
\end{equation}
the Euclidean Feynman propagator is then given by 
\begin{equation}
G_E^F(x,x') = - {i\over 8\pi^2} {1\over Z - 1 - i \epsilon} 
\label{EfeynnuhalfZ}
\end{equation}
so that the de Sitter invariance of the Euclidean vacuum is manifest.  The massless, conformally coupled $\alpha$ propagator is 
\begin{equation}
G_\alpha^F(x,x') = - {iA_\alpha\over 8\pi^2} {1\over Z - 1 - i \epsilon} 
+ {iB_\alpha\over 8\pi^2} {1\over Z + 1 - i \epsilon} . 
\label{AfeynnuhalfZ}
\end{equation}
The second term has resulted from the extra point source at the antipode.  Although it resulted from a non-local interaction in the generating functional, $J(x_A)\Phi(x)$, it has resulted in a completely de Sitter invariant propagator.

\section{Renormalization}
\label{renormalize}

In its original form, the $\alpha$-vacuum leads to non-renormalizable divergences in the perturbative corrections to processes in an interacting theory.  The origin of these divergences is understood by considering the ultraviolet (UV) behavior of Wightman functions which become directly proportional to phases, $G_p^>(\eta_1,\eta_2) \propto e^{-ip(\eta_1-\eta_2)}$ as $p\to\infty$.  An $\alpha$ propagator based on Eq.~(\ref{badAdef}) has all four possible types of phases appearing from the $\alpha$ Wightman functions
\begin{eqnarray}
&&iN_\alpha^2 {\eta_1\eta_2\over 2p} \Bigl[ 
e^{-ip(\eta_1-\eta_2)} + e^{\alpha+\alpha^*} e^{ip(\eta_1-\eta_2)} 
\nonumber \\
&&\qquad
- i e^\alpha e^{-i\pi\nu} e^{ip(\eta_1+\eta_2)} 
+ i e^{\alpha^*} e^{i\pi\nu} e^{-ip(\eta_1+\eta_2)} 
\Bigr] . \qquad
\label{UVbadWight}
\end{eqnarray}
When we have a loop correction, it is possible for an anti-coherence of the phases among different loop propagators so that for large values of the loop momentum, there will be terms in which all of the loop momentum dependence in the exponents cancels.  A given loop correction will then diverge in the UV if the number of powers of the loop momentum in the numerator is greater or equal to the number of loop propagators which scale as $p^{-1}$, as seen in Eq.~(\ref{UVbadWight}).  In \cite{fate} it is shown that this class of divergences cannot be absorbed by a local counterterm of the same form as those already present in the theory.   

The effect of including two sources in the generating functional is to remove some of the terms from the $\alpha$ propagator so that this phase cancellation no longer occurs.  For a local set of interactions, the loop propagators reduce to Euclidean propagators as shown in Eq.~(\ref{JtJtG}).  Then we essentially only have Euclidean Wightman functions
\begin{equation}
G_p^>(\eta_1,\eta_2) \to
{i\eta_1\eta_2\over 2p} 
e^{-ip(\eta_1-\eta_2)} 
\label{UVEuclid}
\end{equation}
in the UV and the time-ordering prevents any constructive phase interference among the loop propagators.  It is still necessary to show that any other divergences, such as logarithmic divergences in loops containing only two propagators, can be renormalized since the external legs are $\alpha$-dependent \cite{fate}.

The effect of restricting to only local counterterms for the double source generating functional is actually more constraining than necessary if we only insist on obtaining a causal, renormalizable theory.  From the perspective of Eq.~(\ref{UVbadWight}) it is the interference of the first or third term in one propagator and second or fourth terms in another, respectively, which leads to a phase cancellation.  It should be necessary to remove only half the terms to eliminate this possibility.  Thus we could consider as well the theory with antipodal non-local interactions as in Eq.~(\ref{specialint}) which contains extra terms in the propagators, as $p\to\infty$, 
\begin{equation}
\tilde G_p^>(\eta_1,\eta_2)
\to
{i\eta_1\eta_2\over 2p} \Bigl[ 
A_\alpha e^{-ip(\eta_1-\eta_2)} 
- i B_\alpha e^{-i\pi\nu} e^{ip(\eta_1+\eta_2)} 
\Bigr] 
\label{UVGtilde}
\end{equation}
where we have defined 
\begin{equation}
\tilde G_p^{>(<)}(\eta_1,\eta_2) \equiv A_\alpha\, G_p^{>(<)}(\eta_1,\eta_2) + B_\alpha\, G_p^>(-\eta_1,\eta_2) . 
\label{Gtildedef}
\end{equation}

\begin{figure}[!tbp]
\includegraphics{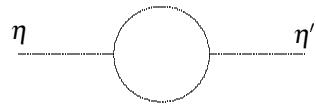}
\caption{The one loop self-energy correction in a $\Phi^3$ theory.
\label{oneloop}}
\end{figure}

In this section we calculate the first non-trivial correction to the propagator, $\Sigma_k^{(1)}(\eta,\eta')$, where
\begin{eqnarray}
&&\!\!\!\!\!\!\!\!
\langle\alpha |\, T_\alpha \bigl( \Phi(x) \Phi(x') \bigr) \, |\alpha\rangle 
\label{Sigmadef} \\
&&= - i G_\alpha^F(x,x') 
- i \int {d^3\vec k\over(2\pi)^3}\, e^{i\vec k\cdot(\vec x-\vec x')} 
\Sigma_k^{(1)}(\eta,\eta') 
+ \cdots 
\nonumber
\end{eqnarray}
for both of these cases to show that both can be renormalized by the appropriate counterterms.  To analyze the self-energy correction in the simplest setting, we shall examine a theory with a cubic interaction since then $\Sigma_k^{(1)}$ corresponds to a one loop graph, shown in Fig.~\ref{oneloop}.

\subsection{Local interactions}

We begin with a cubic local interaction with the associated counterterms,
\begin{equation}
- {\cal L}_I = \rho \Phi(x) 
+ {\textstyle{1\over 2}} \delta m^2 \Phi^2(x) 
+ {\textstyle{1\over 3!}}\lambda \Phi^3(x) .
\label{Lintdef}
\end{equation}
The new linear term has been added since the cubic term explicitly breaks the $\Phi(x)\to -\Phi(x)$ symmetry which was present in the free Lagrangian.  The correct choice of $\rho$ cancels all the graphs in perturbation theory which contain a tadpole subgraph.  At order $\lambda^2$, the mass counterterm is needed to remove a logarithmic divergence produced in the self-energy diagram.

Expanding the Schwinger-Keldysh generating functional given in Eq.~(\ref{SKgenfunc}) and Eq.~(\ref{genfuncJJ4}) to order $\lambda^2$ in the cubic interaction yields
\begin{eqnarray}
&&\!\!\!\!\!\!\!\!\!\!\!\!
\Sigma_k^{(1)}(\eta,\eta') 
\label{Sigmaone} \\
&=&  -i {\lambda^2\over 16\pi^3} \int_{\eta_0}^0 {d\eta_1\over\eta_1^4} 
\int_{\eta_0}^0 {d\eta_2\over\eta_2^4} \int d^3\vec p\, 
\nonumber \\
&&\bigl[ 
G_k^{++}(\eta,\eta_1) G_p^{++}(\eta_1,\eta_2) 
G_{|\vec p-\vec k|}^{++}(\eta_1,\eta_2) G_k^{++}(\eta_2,\eta') 
\nonumber \\
&&
- G_k^{++}(\eta,\eta_1) G_p^{+-}(\eta_1,\eta_2) 
G_{|\vec p-\vec k|}^{+-}(\eta_1,\eta_2) G_k^{-+}(\eta_2,\eta') 
\nonumber \\
&&
- G_k^{+-}(\eta,\eta_1) G_p^{-+}(\eta_1,\eta_2) 
G_{|\vec p-\vec k|}^{-+}(\eta_1,\eta_2) G_k^{++}(\eta_2,\eta') 
\nonumber \\
&&
+ G_k^{+-}(\eta,\eta_1) G_p^{--}(\eta_1,\eta_2) 
G_{|\vec p-\vec k|}^{--}(\eta_1,\eta_2) G_k^{-+}(\eta_2,\eta') 
\bigr] .
\nonumber
\end{eqnarray}
All other diagrams at this order either contain a disconnected vacuum to vacuum subgraph, which is cancelled by the normalization Eq.~(\ref{genvactovac}), or contain a tadpole subgraph, which is cancelled by a corresponding graphs with the tadpole replaced by an insertion of the $\rho\Phi(x)$ counterterm provided
\begin{equation}
\rho = {i\lambda\over 16\pi^3} \int d^3\vec p\, G^>_p(\eta_1,\eta_1) . 
\label{notadpole}
\end{equation}
This counterterm is constant since it can be rewritten as 
\begin{equation}
\rho = {\lambda\over 16\pi} \int_0^{\infty} d\xi\, 
\xi^2 \bigl| H^{(2)}_\nu(\xi) \bigr|^2 . 
\label{notadpoleconst}
\end{equation}

The propagators which contain an external point, $G_k^{+\pm}(\eta,\eta_1)$ and $G_k^{\pm +}(\eta_2,\eta')$, correspond to mixed propagators while the loop propagators are purely Euclidean.  Despite the appearance of an antipodal conformal time in these mixed propagators, the theory remains causal since no information can propagate from times later than $\eta$ or $\eta'$.  This property becomes evident when we expand the propagators in terms of the Wightman functions, using Eqs.~(\ref{JtJtG})--(\ref{JJtG}),
\begin{widetext}
\begin{eqnarray}
\Sigma_k^{(1)}(\eta,\eta') 
&=&  -i {\lambda^2\over 16\pi^3} \int_{\eta_0}^0 {d\eta_1\over\eta_1^4} 
\int_{\eta_0}^0 {d\eta_2\over\eta_2^4} \int d^3\vec p 
\label{SElocal} \\
&& \Bigl\{ a_\alpha^2 \Bigl[
\Theta(\eta-\eta_1) \Theta(\eta_1-\eta_2) \Theta(\eta'-\eta_2) \, 
(G_k^>-G_k^<) (G_p^>G_{|\vec p-\vec k|}^>G_k^< - G_p^<G_{|\vec p-\vec k|}^<G_k^>)
\nonumber \\
&&\quad\ 
+ \Theta(\eta-\eta_1) \Theta(\eta_2-\eta_1) \Theta(\eta'-\eta_2) \, 
(G_k^<G_p^>G_{|\vec p-\vec k|}^> - G_k^>G_p^<G_{|\vec p-\vec k|}^<) (G_k^>-G_k^<) 
\nonumber \\
&&\quad\ 
+ \Theta(\eta-\eta_1) \Theta(\eta_1-\eta_2) \Theta(\eta_2-\eta') \, 
(G_k^>-G_k^<) (G_p^>G_{|\vec p-\vec k|}^>-G_p^<G_{|\vec p-\vec k|}^<) G_k^> 
\nonumber \\
&&\quad\ 
+ \Theta(\eta_1-\eta) \Theta(\eta_2-\eta_1) \Theta(\eta'-\eta_2) \, 
G_k^< (G_p^>G_{|\vec p-\vec k|}^>-G_p^<G_{|\vec p-\vec k|}^<) (G_k^>-G_k^<)
\Bigr]
\nonumber \\
&&\  
+ a_\alpha b_\alpha \Bigl[
\Theta(\eta-\eta_1) \Theta(\eta_1-\eta_2) \, 
(G_k^>-G_k^<) (G_p^>G_{|\vec p-\vec k|}^>-G_p^<G_{|\vec p-\vec k|}^<)
G_k^>(-\eta_2,\eta')
\nonumber \\
&&\quad\ 
+ \Theta(\eta_2-\eta_1) \Theta(\eta'-\eta_2) \, 
G_k^>(-\eta,\eta_1) (G_p^>G_{|\vec p-\vec k|}^>-G_p^<G_{|\vec p-\vec k|}^<) (G_k^>-G_k^<) 
\Bigr]
\Bigr\} . 
\nonumber 
\end{eqnarray}
\end{widetext}

To establish the renormalizability of this theory to one loop order, we shall study the $\nu={1\over 2}$ limit since in this case the Wightman functions reduce to an easily integrated form.  Also for purposes of illustration we set $\eta=\eta'$ to simplify the $\Theta$-function structure of the self-energy although the value for $\delta m^2$ chosen cancels the divergence in when $\eta\not = \eta'$ as well.  The loop integral for the $\nu={1\over 2}$ Wightman functions is finite,
\begin{equation}
\int d^3\vec p\, G_p^>(\eta_1,\eta_2) G_{|\vec p-\vec k|}^>(\eta_1,\eta_2) 
= {i\pi\over 2} {(\eta_1\eta_2)^2\over\eta_1-\eta_2} e^{-ik(\eta_1-\eta_2)} , 
\label{Eloop}
\end{equation}
except when the times $\eta_1$ and $\eta_2$ coincide.  We shall regulate this divergence as in \cite{fate} by continuing the number of time dimensions to $1+\epsilon$ in the $\eta_2$ integration to extract the $\epsilon\to 0$ pole,
\begin{equation}
\int_{\eta_0}^{\eta_1} d\eta_2\, {e^{2iq\eta_2}\over \eta_2(\eta_1-\eta_2)} 
= {e^{2iq\eta_1}\over\eta_1} {1\over\epsilon} + {\rm finite} . 
\label{regulate}
\end{equation}
The divergent part of the self-energy graph is then 
\begin{widetext}
\begin{eqnarray}
\Sigma_k^{(1)}(\eta,\eta) 
&=& {1\over\epsilon} {\lambda^2\over 16\pi^2} \int_{\eta_0}^\eta {d\eta_1\over\eta_1^4} 
\biggl\{ a_\alpha^2 \bigl[
G_k^>(\eta,\eta_1) G_k^<(\eta_1,\eta) - G_k^<(\eta,\eta_1) G_k^>(\eta_1,\eta)
\bigr]
\label{SElocalpole} \\
&&\qquad\qquad\qquad
+\ 2 a_\alpha b_\alpha 
\bigl[ G_k^>(\eta,\eta_1) - G_k^<(\eta,\eta_1) \bigr]
G_k^>(-\eta_1,\eta)
\biggr\}
\nonumber \\
&&+\ {\rm finite} .
\nonumber
\end{eqnarray}
The corresponding contribution from the mass counterterm at leading order, $\Sigma_k^{\delta m^2}(\eta,\eta)$, is of the same form,
\begin{eqnarray}
\Sigma_k^{\delta m^2}(\eta,\eta) 
&=& - \delta m^2 \int_{\eta_0}^\eta {d\eta_1\over\eta_1^4} 
\biggl\{ a_\alpha^2 \bigl[
G_k^>(\eta,\eta_1) G_k^<(\eta_1,\eta) - G_k^<(\eta,\eta_1) G_k^>(\eta_1,\eta)
\bigr]
\label{dmlocal} \\
&&\qquad\qquad\qquad
+\ 2 a_\alpha b_\alpha 
\bigl[ G_k^>(\eta,\eta_1) - G_k^<(\eta,\eta_1) \bigr]
G_k^>(-\eta_1,\eta)
\biggr\} , 
\nonumber
\end{eqnarray}
\end{widetext}
so that the divergence is cancelled when
\begin{equation}
\delta m^2 = {\lambda^2\over 16\pi^2} {1\over\epsilon} . 
\label{dmlocalfix}
\end{equation}

Note that only local counterterms were necessary to remove the divergences in the self-energy diagram.

\subsection{Special antipodal interactions}

The local, cubic interaction Lagrangian can be generalized in de Sitter space to a Lagrangian with non-local terms that contain the antipode, 
\begin{equation}
- {\cal L}_I = \rho \widetilde\Phi(x)
+ {\textstyle{1\over 2}} \delta m^2 \widetilde\Phi(x)^2  
+ {\textstyle{1\over 3!}}\lambda \widetilde\Phi(x)^3 .
\label{LNLintdef}
\end{equation}
where $\widetilde\Phi(x)$ is given by Eq.~(\ref{specialfield}).  Note that because the cubic term generates a diagram in which all propagators are pure $\alpha$-vacuum propagators of the form given by Eq.~(\ref{alphaprop}), the mass counterterm needed to remove the logarithmic divergence seen above will also need to be of the form $\widetilde\Phi^2(x)$ since a purely local mass counterterm would have produced a product of mixed propagators as in Eq.~(\ref{dmlocal}).  As before, diagrams with a tadpole subgraph are removed by choosing 
\begin{eqnarray}
\rho 
&=& {i\lambda\over 16\pi^3} \int d^3\vec p\, \tilde G^>_p(\eta_1,\eta_1)  
\label{NLnotadpole} \\
&=& 
{\lambda\over 16\pi} \int_0^{\infty} d\xi\, 
\xi^2 \Bigl[ 
A_\alpha \bigl| H^{(2)}_\nu(\xi) \bigr|^2 
\nonumber \\
&&\qquad\qquad\qquad
+ B_\alpha e^{-i\pi\left( \nu - {1\over 2} \right)} 
\bigl( H^{(2)}_\nu(\xi) \bigr)^2 \Bigr] . 
\nonumber
\end{eqnarray}

Once the tadpole graphs have been removed, the only remaining graph at order $\lambda^2$ is the self-energy graph shown in Fig.~\ref{oneloop} which contributes
\begin{widetext}
\begin{eqnarray}
\Sigma_k^{(1)}(\eta,\eta') 
&=&  -i {\lambda^2\over 16\pi^3} \int_{\eta_0}^0 {d\eta_1\over\eta_1^4} 
\int_{\eta_0}^0 {d\eta_2\over\eta_2^4} \int d^3\vec p 
\label{SEnonlocal} \\
&& \Bigl\{ A_\alpha^2 \Bigl[
\Theta(\eta-\eta_1) \Theta(\eta_1-\eta_2) \Theta(\eta'-\eta_2) \, 
(G_k^>-G_k^<) (\tilde G_p^>\tilde G_{|\vec p-\vec k|}^>G_k^< - \tilde G_p^< \tilde G_{|\vec p-\vec k|}^< G_k^>)
\nonumber \\
&&\quad\ 
+ \Theta(\eta-\eta_1) \Theta(\eta_2-\eta_1) \Theta(\eta'-\eta_2) \, 
(G_k^<\tilde G_p^>\tilde G_{|\vec p-\vec k|}^> - G_k^> \tilde G_p^< \tilde G_{|\vec p-\vec k|}^<) (G_k^>-G_k^<) 
\nonumber \\
&&\quad\ 
+ \Theta(\eta-\eta_1) \Theta(\eta_1-\eta_2) \Theta(\eta_2-\eta') \, 
(G_k^>-G_k^<) (\tilde G_p^>\tilde G_{|\vec p-\vec k|}^>-\tilde G_p^<\tilde G_{|\vec p-\vec k|}^<) G_k^> 
\nonumber \\
&&\quad\ 
+ \Theta(\eta_1-\eta) \Theta(\eta_2-\eta_1) \Theta(\eta'-\eta_2) \, 
G_k^< (\tilde G_p^>\tilde G_{|\vec p-\vec k|}^>-\tilde G_p^<\tilde G_{|\vec p-\vec k|}^<) (G_k^>-G_k^<)
\Bigr]
\nonumber \\
&&\  
+ A_\alpha B_\alpha \Bigl[
\Theta(\eta-\eta_1) \Theta(\eta_1-\eta_2) \, 
(G_k^>-G_k^<) 
(\tilde G_p^>\tilde G_{|\vec p-\vec k|}^>-\tilde G_p^< \tilde G_{|\vec p-\vec k|}^<)
G_k^>(-\eta_2,\eta')
\nonumber \\
&&\qquad\ 
+ \Theta(\eta_2-\eta_1) \Theta(\eta'-\eta_2) \, 
G_k^>(-\eta,\eta_1) 
(\tilde G_p^>\tilde G_{|\vec p-\vec k|}^>-\tilde G_p^< \tilde G_{|\vec p-\vec k|}^<) (G_k^>-G_k^<) 
\Bigr]
\Bigr\}
\nonumber
\end{eqnarray}
\end{widetext}
The formal structure of the self-energy contribution is exactly the same as in the previous case and is explicitly causal because of the structure of the $\Theta$-functions.  In addition to containing the coefficients associated with the $\alpha$ propagator, the loop integrals now have a dependence on the antipodes as well.

As before, we evaluate the loop integrals for the $\nu={1\over 2}$ Wightman functions, 
\begin{eqnarray}
&&\!\!\!\!\!\!\!\!\!\!\!\!\!\!\!\!\!\!\!\!\!\!\!\!
\int d^3\vec p\, \tilde G_p^>(\eta_1,\eta_2) \tilde G_{|\vec p-\vec k|}^>(\eta_1,\eta_2) 
\nonumber \\
&=& {i\pi A_\alpha^2\over 2} {(\eta_1\eta_2)^2\over\eta_1-\eta_2} e^{-ik(\eta_1-\eta_2)}
\nonumber \\
&&
- {i\pi B_\alpha^2\over 2} {(\eta_1\eta_2)^2\over\eta_1+\eta_2} e^{ik(\eta_1+\eta_2)}
\nonumber \\
&&
+ {i\pi A_\alpha B_\alpha\over k} \eta_1\eta_2 e^{ik\eta_2} \sin(k\eta_1) . 
\label{Aloop}
\end{eqnarray}
The only possible divergence occurs in the first term when $\eta_2\to\eta_1$ so that the presence of the additional antipodal terms in the loop propagator has not worsened the renormalizability of the theory, unlike the case of the loop integral of a single-source $\alpha$ propagator \cite{fate,einhorn1,banks}.  The divergent part of $\Sigma_k^{(1)}(\eta,\eta')$, evaluated for $\nu={1\over 2}$, is of the same form as that for the local interactions in Eq.~(\ref{SElocalpole}) except with an overall factor of $A_\alpha^2$ and with $a_\alpha\to A_\alpha$ and $b_\alpha\to B_\alpha$.  The non-local mass counterterm, evaluated for a general $\nu$, is as in Eq.~(\ref{dmlocal}) with $a_\alpha\to A_\alpha$ and $b_\alpha\to B_\alpha$ so that the divergent pole can be removed by choosing 
\begin{equation}
\delta m^2 = {\lambda^2\over 16\pi^2} {A_\alpha^2\over\epsilon} . 
\label{dmnonlocalfix}
\end{equation}
Note that the appearance of $A_\alpha^2$ indicates that the counterterm is explicitly $\alpha$-dependent.

\section{General antipodal interactions}
\label{antipodes}

The examples in the preceding section illustrate that the set of renormalizable field theories in de Sitter space can have a much richer structure than in flat space.  When the double source generating functional is used, any theory with a Lagrangian density of the form
\begin{eqnarray}
{\cal L} &=& 
{\textstyle{1\over 2}} g^{\mu\nu} \partial_\mu \Phi(x) \partial_\nu \Phi(x) 
- {\textstyle{1\over 2}} m^2 \Phi^2(x) 
\nonumber \\
&&
- \sum c_{p,q} \Phi^p(x) \Phi^q(x_A) , 
\label{genint}
\end{eqnarray}
should be renormalizable for $p+q\le 4$ in the following sense:  for an interaction $\Phi^p(x)\Phi^q(x_A)$, divergences in perturbative corrections to any process can be removed by including the most general set of counterterms, $\Phi^{p'}(x)\Phi^{q'}(x_A)$ with $p'+q'\le p+q$.  Note that in a strict sense, some of the counterterms will not be of the same form as the terms in the bare Lagrangian.  In the theory with a local $\Phi^3(x)$ interaction, a mass counterterm was sufficient while the $\widetilde\Phi^3(x)$ theory required a $\widetilde\Phi^2(x)$ counterterm which is not of the same form as the original mass term.  A theory with a more general mixed interaction, such as $\Phi^2(x)\widetilde\Phi(x)$, requires the full set of possible quadratic counterterms---$\Phi^2(x)$, $\Phi(x)\widetilde\Phi(x)$ and $\widetilde\Phi^2(x)$.  

\begin{figure}[!tbp]
\includegraphics{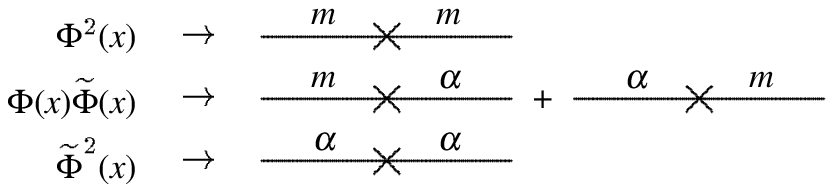}
\caption{The counterterm graphs corresponding to the possible quadratic terms.  The labels $m,\alpha$ indicate whether a propagator is mixed or $\alpha$.
\label{quadterms}}
\end{figure}

In this section we show the renormalizability of self-energy graphs for the case of a general cubic set of interactions, $p+q=3$.  While potentially many graphs, differing in their propagator content, can contribute to a generic self-energy correction, the important feature is that in each case the divergent part of the loop arises from the Euclidean parts of the loop propagators so that up to different overall constant coefficients, the divergent part of the loop integral in the $\eta_2\to\eta_1$ limit is proportional to 
\begin{equation}
\pm {i\pi\over 2} {(\eta_1\eta_2)^2\over\eta_1-\eta_2} e^{\mp ik(\eta_1-\eta_2)} . 
\label{gendivpart}
\end{equation}
The first term on the right side of Eq.~(\ref{Aloop}), for example, has this behavior.  Since the external vertices of the two-point function arise from $-i{\delta\over\delta J(x)}$ derivatives, 
\begin{eqnarray}
&&\!\!\!\!\!\!\!\!\!\!\!\!\!\!\!\!\!\!\!\!\!\!\!\!
\langle\alpha|\, T_\alpha \bigl( \Phi(x) \Phi(x') \bigr)\, |\alpha\rangle 
\nonumber \\
&=& \left[ -i{\textstyle{\delta\over\delta J^+(x)}} \right] 
\left[ -i{\textstyle{\delta\over\delta J^+(x')}} \right] W^\alpha[J^+,J^-] ,
\label{gentwopoint}
\end{eqnarray}
the external legs will always be either mixed or $\alpha$ propagators and can thus be renormalized by one of the quadratic counterterms shown in Fig.~\ref{quadterms}.

To show that the self energy corrections for a set of cubic interactions,
\begin{eqnarray}
{\cal L}_{\rm cubic} &=& {\textstyle{1\over 6}} \lambda_0 \Phi^3(x) 
+ {\textstyle{1\over 6}} \lambda_1 \Phi^2(x) \Phi(x_A) 
\nonumber \\
&&
+ {\textstyle{1\over 6}} \lambda_2 \Phi(x) \Phi^2(x_A) 
+ {\textstyle{1\over 6}} \lambda_3 \Phi^3(x_A) , 
\label{gencube}
\end{eqnarray}
can be canceled by the graphs in Fig.~\ref{quadterms}, it is useful to rewrite the interactions in the form of products of local and special antipodal operators,
\begin{eqnarray}
{\cal L}_{\rm cubic} &=& {\textstyle{1\over 6}} \tilde\lambda_0 \Phi^3(x) 
+ {\textstyle{1\over 6}} \tilde\lambda_1 \Phi^2(x) \widetilde\Phi(x) 
\nonumber \\
&&
+ {\textstyle{1\over 6}} \tilde\lambda_2 \Phi(x) \widetilde\Phi^2(x) 
+ {\textstyle{1\over 6}} \tilde\lambda_3 \widetilde\Phi^3(x) . 
\label{gencubet}
\end{eqnarray}
Once in this form, the interacting theory generating functional becomes 
\begin{widetext}
\begin{equation}
W^\alpha[J^+,J^-] = {\cal N} e^{{i\over 6}\int d^4x\, \sqrt{-g} 
\sum_{j=0}^{3} \tilde\lambda_j \left\{ 
\left[ -i {\delta\over\delta{\cal J}^+(x)} \right]^j 
\left[ -i {\delta\over\delta J^+(x)} \right]^{p+q-j}
- \left[ -i {\delta\over\delta{\cal J}^-(x)} \right]^j 
\left[ -i {\delta\over\delta J^-(x)} \right]^{p+q-j} 
\right\} } 
W_0^\alpha[J^+,J^-] . 
\label{gengenfunc}
\end{equation}
\end{widetext}
where the free generating functional is given by Eq.~(\ref{genfuncJJ4}).

\begin{figure}[!tbp]
\includegraphics{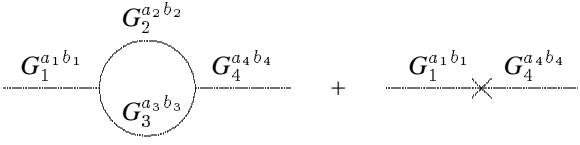}
\caption{A one loop self-energy correction in a general cubic theory with an arbitrary propagator structure.  Note that the external legs are either mixed or $\alpha$ propagators.  The counterterm diagram removes the logarithmic divergence of the loop.
\label{SEgen}}
\end{figure}

We analyze the divergence structure for a completely arbitrary self-energy graph by writing the coefficients of the Euclidean part of the propagator and the antipodal part as ${\cal A}_i$ and ${\cal B}_i$, respectively, where the subscript refers to a particular line in the graph as shown in Fig.~\ref{SEgen}.  These coefficients correspond to those defined in Eqs.~(\ref{countourGreens}--\ref{Acase}).  Once we have integrated over the loop three-momentum and extracted the term which diverges in the limit in which the conformal times of the two internal vertices coincide, the $(\eta, k)$-dependence of the divergent part of the the loop graph in Fig.~\ref{SEgen} is precisely of the same form as counterterm graph shown with the same external legs.

Neglecting constant combinatoric factors, the divergent part of the loop diagram in Fig.~\ref{SEgen} is 
\begin{widetext}
\begin{eqnarray}
\Sigma_k^{({\rm Fig.}~\ref{SEgen})}(\eta,\eta) 
&\propto& {1\over\epsilon} {\lambda_j\lambda_{j'}\over 16\pi^2} {\cal A}_2 {\cal A}_3 \int_{\eta_0}^\eta {d\eta_1\over\eta_1^4} 
\biggl\{ {\cal A}_1 {\cal A}_4 \bigl[
G_k^>(\eta,\eta_1) G_k^<(\eta_1,\eta) 
- G_k^<(\eta,\eta_1) G_k^>(\eta_1,\eta)
\bigr]
\nonumber \\
&&\qquad\qquad\qquad\qquad
+\ [ {\cal A}_1 {\cal B}_4 + {\cal B}_1 {\cal A}_4 ]
\bigl[ G_k^>(\eta,\eta_1) 
- G_k^<(\eta,\eta_1) \bigr]
G_k^>(-\eta_1,\eta)
\biggr\}
\nonumber \\
&&+\ {\rm finite} ,
\label{SEgenpole}
\end{eqnarray}
which we have written for the $\eta=\eta'$ case for simplicity.  Here we allowed for the vertices to have arisen from different interaction terms with couplings $\lambda_j$ and $\lambda_{j'}$.  Since this graph is proportional to ${\cal A}_2 {\cal A}_3$, the origin of the divergence is solely from the Euclidean part of the propagators in the loop.  Comparing this structure of this term with that of the appropriate counterterm, 
\begin{eqnarray}
\Sigma_k^{\delta m^2}(\eta,\eta) 
&=& - \delta m^2 \int_{\eta_0}^\eta {d\eta_1\over\eta_1^4} 
\biggl\{ {\cal A}_1 {\cal A}_4 \bigl[
G_k^>(\eta,\eta_1) G_k^<(\eta_1,\eta) 
- G_k^<(\eta,\eta_1) G_k^>(\eta_1,\eta)
\bigr]
\nonumber \\
&&\qquad\qquad\qquad
+\ [ {\cal A}_1 {\cal B}_4 + {\cal B}_1 {\cal A}_4 ]
\bigl[ G_k^>(\eta,\eta_1) 
- G_k^<(\eta,\eta_1) \bigr]
G_k^>(-\eta_1,\eta)
\biggr\} , \quad
\label{dmgen} 
\end{eqnarray}
\end{widetext}
we discover that the proper choice of $\delta m^2$ removes the divergence.

\section{Conclusions}
\label{conclude}

de Sitter space appears to admit a much larger set of renormalizable field theories, based on the invariant $\alpha$-states, than flat space.  For any point $x$ in de Sitter space there exists a unique antipode $x_A$ and the invariant distance between $x_A$ and another point $y$ depends only on the invariant distance between $x$ and $y$.  This property allows for the construction of a new propagator, with a point source and a source at its antipode, in terms of which theories with antipodal interactions of the form $\Phi^p(x)\Phi^q(x_A)$ can be renormalized and are causal.  The construction of this theory depended critically on the global geometry of de Sitter space.  

While the theory is manifestly non-local, this non-locality only appears in a global form; an arbitrary point in de Sitter space is always causally disconnected from its antipode.  It would be useful to examine these theories further to determine whether this non-locality ever produces a more severe breakdown of locality, for example between generic points within the same causal patch.  We have also only checked the renormalizability of the theories to one loop order and only for a loop containing two propagators, although processes with a similar loop structure such as the first corrections to the vertex in a theory with a quartic coupling should proceed similarly.  It would be interesting to check whether this renormalizability persists to all orders and to see whether any new features appear in loops with three propagators, such as the vertex corrections in the theory with cubic interactions.

Finally, these states should be simpler to analyze in a cosmological setting than the original $\alpha$-vacuum with a single source propagator.  Even in a classical background, an interacting field theory in the original $\alpha$-vacuum requires a cutoff.  Although we might still need to modify the $\alpha$-states described in this article when the physical momentum approaches the Planck mass \cite{lowe2,einhorn2}, they can be analyzed in a classical gravitational background even without a cutoff.  Since these de Sitter invariant states are renormalizable, it becomes possible to study in principle whether any is preferred---whether a universe that starts in a general $\alpha$-state tends to evolve through the presence of interactions towards a particular value of $\alpha$.  For example, if the interactions allow the state to thermalize at a rate determined by the interaction strength $\lambda$, then if the thermalization proceeds sufficiently rapidly compared over a Hubble time, $H^{-1}$, then a Bunch-Davies vacuum could naturally result.  This process should be able to be addressed using the resummation approach of \cite{dan}.

\begin{acknowledgments}

This work was supported in part by DOE grant DE-FG03-91-ER40682.  

\end{acknowledgments}

\end{document}